\documentstyle[prd,aps,epsfig]{revtex}
\def\be{\begin{equation}}
\def\ee{\end{equation}}
\begin{document}
\draft
\title{Ultraviolet cut off and Bosonic Dominance}
 
\author{Musongela Lubo \footnote{E-mail muso@ictp.trieste.it} \\
        The Abdus Salam International Centre for Theoretical Physics P.O.Box
586\\
34100 Trieste, Italy}
  
\maketitle
\begin{abstract}
 We rederive the thermodynamical properties of a non interacting gas in the 
 presence of a minimal uncertainty in length. Apart from the phase space
 measure which is modified due to a  change of the Heisenberg uncertainty
 relations, the presence of an ultraviolet cut-off plays a tremendous role.
 
 The theory admits  an intrinsic temperature above which the fermion 
 contribution to energy density, pressure and entropy is negligible.
\end{abstract}
\section{Introduction}
\par Modern physics is an edifice in which every stone is tightly linked
to the others. A slight modification in one area may produce
important changes in  different fields. Quantum mechanics starts
with the commutation relations. Once they have been fixed, using a
representation of the operator algebra we can in principle solve
the Schr\"odinger equation whose solutions potentially contain all
the physics of the (non relativistic) system studied. The
correspondence principle which states the link between the
commutator of two variables of the phase space and their classical
Poisson bracket is one of the basic axioms of quantum mechanics.
It is not deduced from another assumption and cannot be judged in
an isolated manner: only the whole theory can be confronted to
experience through its predictions.

\par Some authors have investigated the consequences of the alterations of
these commutation relations on the observables of some physical
systems. In particular, some deformations
of the canonical commutators studied by Kempf-Mangano-Mann(K.M.M)
induce a minimal uncertainty in position(or momentum) in a
very simple way, providing a toy model with manifest non locality \cite{LABEL1}.
The implications of some of these quantum structures have been
studied for the harmonic oscillator \cite{LABEL1,LABEL2,LABEL3} and the
hydrogen atom \cite{LABEL5}. The transplanckian problem occurring
in the usual description of the Hawking mechanism of black hole
evaporation has also been addressed in this framework, for the
Schwarzschild and the Banados-Teitelboim-Zanelli(BTZ) solutions
\cite{LABEL32,LABELl2}. These studies have extended the work on the
entropy of the black hole \cite{Bek} and 
Hawking radiation \cite{Haw} in theories with modified dispersion relations
\cite{Un2,Un1,Jac,co,brgr}

\par One of the purposes of this article is a re-investigation of the modifications
these models induce, not in the characteristics of a single
particle, but in the behavior of a macroscopic system. The modern
presentation of thermodynamics basically relies on statistical
physics. According to the system under study(isolated, closed or
opened), one uses an ensemble(micro canonical, canonical or grand
canonical) and the corresponding potential(entropy, free energy,
grand potential) to derive thermodynamic quantities(pressure,
specific heat, chemical potential, etc).
\par  The thermodynamic
potentials are related to the derivatives of the partition
function which itself is an average(of a quantity which depends on
the ensemble used) on the phase space. To define a measure on the
phase space, one needs to know the extension of the fundamentals
cells. In the ``classical''statistical physics, the Heisenberg
uncertainty relations are used to show that this volume is  the
cube of the Planck constant and the undiscernability of particles
justifies the Gibbs factor.
 This can be  inferred from quantum statistical physics
in which the sum giving the partition function  is made on
a discrete set of states.
\par  If one modifies the
commutators, one changes  the Heisenberg
uncertainty relations. The measure on the phase space is no more
the same; this results in new partition functions and
consequently different thermodynamical behaviors. From the quantum point of
view, the energy spectrum of systems are modified by the change in
the commutation relations.

The thermodynamics of some models displaying a minimal uncertainty in length
has been analyzed before \cite{mons,rama,lay}. We shall give here better
approximations and correct some sign errors. The implications for the early
universe have been at the center of many works among which \cite{ni1,nifin}.
The standard big bang in a universe in which an ultraviolet cut off appears 
in a toy model exhibiting a modified dispersion relation has similarly been 
analyzed \cite{LABEL21}. The difference between
the two approaches  relies in the fact that 
the  dispersion  relations are different; in the first models, they come from 
an assumption on the structure of the quantum phase space.

In these works, the equation of state used  for radiation was  
obtained considering bosons. In the usual case the contributions of
fermions to energy, pressure and entropy are simply the seven eights of the ones
 corresponding to  bosons. This  renders the equation of state insensitive to the
ratio between fermionic  and bosonic degrees of freedom. We show this
to be drastically changed in the new framework.

Many studies have been devoted to cosmological perturbations in transplanckian
physics \cite{mb,bm,iap,nie,np,tanaka,kempf,gree1,EGK01,mbk,shanka1,shanka2}. The 
considerations we develop here may, in this context, be seen as  relevant
only in the pre-inflationary era. However, the phenomenological bounds 
\cite{LABEL5} are much lower than the Planck scale. If one adopts a less
restrictive point of view, then the new scale may  generate some seizable 
effects.   
 
\par The article is organized as follows. In the second section we give
a very brief survey of the two models we will study. They possess a minimal 
length uncertainty and so the quasi-position representation plays a crucial 
role. The third section is devoted to the study of the non interacting gas. We
obtain its equation of state, entropy, etc. We use its specific heat to
set a bound on the deformation parameter and find it to be in agreement with previous 
ones. In the fourth section we address the question of the black-body radiation
in the new context and find sensible differences between fermions and bosons
at very high temperatures, like for the ultra relativistic gas.

\section{Deformed theories}
 
As shown by KMM \cite{LABEL1}, the quantum mechanical theory defined by the commutation 
relation
\be
\label{eq1}
[ \hat x, \hat p] = i \hbar (1 + \beta {\hat p}^2 )
\ee
is endowed with a minimal length uncertainty $\Delta x_{min} = \hbar \sqrt{\beta}$.
The presence of this minimal length uncertainty implies that no
position representation exists. The concept which proves to be the
closest to it is the quasi-position representation in which the
operators are non local:
\be
\label{eq2}
 \hat p = \frac{1}{\sqrt{\beta}} \tan{ \left( i \hbar \sqrt{\beta} 
 \partial_{\xi} \right)} \quad , \quad \hat x = \xi +
 \hbar \sqrt{\beta}  \tan{ \left( i \hbar \sqrt{\beta} 
 \partial_{\xi} \right)}  \quad .
\ee

\par The most obvious extension to a three dimensional space 
is obtained by taking the tensorial product of three such
copies. It will be referred to as the $A_1$ model. It
has translational invariance but lacks the symmetry under
rotations. Another extension  preserves  rotational and translational
symmetry; it will be
referred to as the $A_2$ model. Its commutation relations are
\be
\label{eq3}
[ \hat x_j, \hat p_k ] = i \hbar \left( f({\hat p}^2) \delta_{j,k} + 
 g({\hat p}^2) \hat p_j \hat p_k \right) 
 \quad , \quad g(p^2) = \beta \quad , \quad
 f(p^2) = \frac{\beta p^2}{-1+\sqrt{1+ 2 \beta p^2}} \quad .
\ee
We will need the form of the momentum operators in this model:
\be
\label{eq4}
p_k = - i \hbar \sum_{r=0}^{\infty}  \left( \frac{\hbar^2 \beta}{2} 
\Delta \right)^r \frac{\partial}{\partial \xi_{k}} \quad , \quad {\rm where} 
\quad
\Delta = \sum_{l=1}^3  \frac{\partial^2}{\partial \xi_{l}^2} \quad .
\ee

\section{The  non interacting gas in deformed theories}

Introducing the momentum scale $ \beta$ , it is straightforward
that with the Boltzmann constant $k$ , the light velocity $c$ and
the mass $m$ of any particle, one can construct on purely
dimensional grounds the characteristic temperatures 
\be
\label{eq5}
 T_c =
\frac{1}{\beta m k} \quad , \quad T_{cr} =  \frac{c}{k
\sqrt{\beta}} \quad . \ee The first is ``non relativistic'',
particle dependent while the second is ``relativistic''($c$
dependent) and universal. We are interested in what happens at and
above these temperatures.

\par The constant $\beta $ which appear in the KMM algebra is a free parameter.
 What numerical value can it assume? It was suggested \cite{LABEL015}
that the minimal length uncertainty $ \hbar \sqrt{\beta} $ should
 be of the order of the Planck scale. We shall adopt a less restrictive point of view here.
 The only constraint is that  the deformed commutator  
 should not lead to contradictions with the predictions of the orthodox 
  theory which have been observed experimentally. Our attitude is inspired
 by recent works which have shown that a new physics may take place  well before
 the Planck scale \cite{LABEL016}. Taking the most  stringent phenomenological 
constraint $ \hbar \sqrt{\beta} \leq
10^{-16} m \cite{LABEL5}$, one finds a lower bound to the
characteristi crelativistic  temperature : $ T_{cr} \geq 10^{13}
°K$. If one assumes the minimal length uncertainty to be of the
order of the Planck scale, then $ T_{cr} = 10^{16}   °K $.

We now briefly summarize the formula of statistical physics we will need.
 In the  usual theory, a system which is in contact with a
 large heat reservoir and doesn't exchange particles with the surroundings has to be studied in the canonical ensemble
\cite{LABEL013},\cite{LABEL014}. Its equilibrium state will be
described by a fixed temperature and a fixed particle number while
its energy will fluctuate around a mean value. Strictly speaking,
for such a system the particle number $ N $ is fixed once and for
all. But, one knows that when phase transitions are not present,
the description given by the canonical and the grand canonical
ensembles are very close. This will be used to compute the
chemical potential in the canonical ensemble where the
calculations are easier.

The most important quantity will be the canonical partition function
 $Z(T,V,N)$ which is defined in terms of
the Hamiltonian  operator $\hat H$ by the equation 
\be
\label{eq6}
 \label{ther6} Z(T,V,N) =
Tr \exp{ \left( - \frac{\hat H}{k T} \right)} \quad . \ee 
The free energy is related to the
partition function by 
\be 
\label{eq7}
 F(T,V,N) = -k T \ln Z \quad .
\ee 
In these variables  the pressure $P$, the entropy $S$,
 the chemical potential $\mu$, the internal energy $U$ and the constant volume
specific heat $C_V$   read
 \be
\label{eq8} P = -{\partial F\over{\partial V}} \quad , \quad  S
= -{\partial F\over {\partial T}} \quad , \quad  \mu = {\partial
F\over{\partial N}} \quad , \quad
 U = F + T S \quad , \quad {\rm and} \quad
 C_V = {\partial U\over
{\partial T}} \quad .
 \ee
 
\subsection{The $A_1$ model}

\subsubsection{Low temperatures}

Low temperatures correspond to non relativistic behaviours. Let us study a non interacting non relativistic gas. One has to solve the
Schr\"odinger equation for a particle in a cubic box of length
$L$. This will be done in the quasi position 
representation because of the lack of a position one.
For simplicity, let us first consider a one dimensional system;
one obtains the solution 
\be 
\label{eq9}
 \psi(t,\xi)\div e^{-iEt} \exp \left(\pm
{i\xi\over{\hbar \sqrt{\beta}}} \arctan \sqrt{2m\beta \hbar E}
\right) \quad . 
\ee 
When the periodic boundary condition
$\psi(t,\xi=0)=\psi(t,\xi=L)$ is imposed, one finds the
quantification of energy \be
 \label{eq10}
  E_n = {1\over{2m\beta}} \tan^2
\left( {2\pi  \hbar \sqrt{\beta} n \over L  } \right) \quad ,
 \ee
 (n
being an integer) already obtained in \cite{LABEL1,LABEL32}. This
leads to a cut  off in  order to avoid a non monotonic dispersion relation:
\begin{equation}
\label{eq11} n_{sup} = E \left[ \frac{L}{4 \hbar \sqrt{\beta}}
\right] \quad ,
\end{equation}
$E$ being the integer part function(not to be confused with the energy).

This cut off $n_{sup}$ is necessary in order to
prevent a divergence of the partition function which would take place 
otherwise, due to the periodic nature of the  energy(Eq.(\ref{eq10})).
But this still allows an infinite energy provided that
the  length of the box is fine tuned  in  such a way that the result of its 
division by the minimal length uncertainty is an integer. This would not be
problematic since such an energy would have a vanishing contribution to
the partition function. This conclusion will remain untouched
at high temperatures.

The one particle partition function is given by the formula
\begin{equation}
 \label{eq12}
  Z(T,V,1) =  \sum_{n=0}^{n_{sup}}
e^{-{E_n\over{kT}}}
 =   \sum_{n=0}^{n_{sup}}   \exp{ \left(- \frac{1}{2  \beta m k
T} \tan^2{\left( \hbar \sqrt{\beta} \frac{2 \pi n}{L} \right)
}\right)} \quad .
 \end{equation}
The sum on $n$ can be approximated by an integral on $ dn $ if the
size of the box $L$ is big enough. Introducing the integration
variable $p$ by
\begin{equation}
\label{13} n = \frac{L}{2 \pi \hbar \sqrt{\beta}} \arctan{
\left( \sqrt{\beta}p \right)}
\end{equation} and replacing, as usual in statistical physics, the length $ L $ by an
integral on position, we find \be
 \label{eq14}
 Z(T,V,1) = {1\over h}\int dx \, dp \quad {1\over{1+\beta
p^2}} \quad e^{-p^2/2mkT} \quad . \ee

\par The formula given in Eq.(\ref{eq14}) admits a semi classical
   interpretation. Let us first consider its limiting case
   $\beta = 0$. Classically, the system can  be seen as a
   point evolving in  phase space. The probability for the
   system to be in a configuration in which the first particle is
   in the region $ \vec{q_1} \pm \vec{dq_1}, \vec{p_1} \pm \vec{dp_1}$ , the second particle in
   the region $ \vec{q_2} \pm  \vec{dq_2}, \vec{p_2} \pm  \vec{dp_2}, \cdots $  is proportional to $
   e^{
   - E( \vec{q},\vec{p})} $ and proportional to the number of elementary cells contained in
   the volume of the
   aforementioned region. At the quantum level, the Heisenberg
   uncertainty relation of the usual theory(written in the one dimensional 
   case)
    $ \Delta p_i \Delta q_i \geq \hbar/2  $
   assigns to each elementary cell a volume $  h $. The number of
   such cells contained in the region under consideration is therefore
\be
\label{eq15}
  \prod_{i=1}^N \frac{ d^3 \vec{p_i} d^3 \vec{q_i} }{h^3} \quad .
\ee
One sees that in the new theory, one
can simply keep the usual dispersion relation and  modify the
elementary cell volume. At the quantum level, this appears as a
Jacobian linked to the change of variables($ \vec n \rightarrow
\vec p$). This could be anticipated with a semi classical
reasoning. The new Heisenberg uncertainty relation implies 
 \be
 \label{eq16}
  \Delta x \Delta p \geq \frac{ \hbar}{2} \left( 1+ \beta \langle p^2 
  \rangle \right) \quad .
  \ee
 It assigns to the elementary cells of the phase
space of the new theory a volume $ h ( 1+ \beta p^2) $ which
replaces the usual factor $ h$. From this we could conjecture a
simple recipe when dealing with the semi classical approximation:
it is obtained by keeping the classical dispersion relation but
modifying the measure in a way consistent with the new Heisenberg
uncertainty relation.

However, it should be noted that, in the new theory, the range over which one
integrates is finite, due to the presence of the cut-off which
depends on the volume but not the temperature:
 \be
 \label{eq17}
 p_{sup}=
\frac{t}{\sqrt{\beta}}   \quad , \quad {\rm with} \quad t
=\tan{ \left\lbrace \frac{\pi}{2} \left( \frac{L}{4 \hbar
\sqrt{\beta}} \right)^{-1} E \left( \frac{L}{4 \hbar \sqrt{\beta}}
\right) \right\rbrace}.
 \ee
 Due to the equality
\be
 \label{eq18}
 \lim_{x \rightarrow \infty} \frac{E(x)}{x} = 1 \quad ,
 \ee
the upper bound goes to infinity and the volume of the elementary
cell tends to the usual one as the deformation parameter is sent
to zero. It can be anticipated that, because the integrand of
Eq.(\ref{eq14}) is a rapidly decreasing function, taking the
upper bound to be infinite will not introduce an appreciable error
in most cases.

\par Going to the three dimensional  extension $A_1$ , one has simply
to take the product of the elementary cells in the three
directions. The integral over the positions is obvious; the change
of variable
 $w=\vec p^2/2mkT$, allows one to write the one point partition
 function as
 \begin{eqnarray}
 \label{eq19}
   Z(T,V,1) &=& \frac{V}{\lambda^3} J \quad , \quad {\rm with} \quad 
 J = \left\lbrace  \frac{1}{\sqrt{\pi}} \int_0^{\omega_{sup}} \exp{(-\omega)}
 \omega^{-1/2} \left(1+ 2 \frac{T}{T_c} \omega \right)^{-1}  d\omega \right\rbrace^3
\quad , \nonumber\\
 \lambda &=& \left( \frac{h^2}{2 \pi m k T} \right)^{1/2}   \quad , \quad
\omega_{sup} = \frac{t^2}{2} \frac{T_c}{T} \quad .
 \end{eqnarray}
 
 When $ \beta$ vanishes, the upper
bound equals  infinity and the integral giving $J$ assumes the
value one so that the undeformed theory  is
recovered. The total partition
function $Z$ is found to be related to the usual one (corresponding
to $\beta=0$ and now denoted $Z_{\ast} )$ by the relation \be
 \label{eq20} Z =
Z_{\ast} J^N  \quad . \ee The free energy then becomes \be
 \label{eq21} F =
F_{\ast} - N k T \ln J \quad .
 \ee
The thermodynamical quantities  are affected in the following way
: 
\begin{eqnarray}
 \label{eq22} P &=& P_{\ast} + N k T {1\over J} {\partial
J\over{\partial V}} \quad , \quad
 S = S_{\ast} +
N k \ln J + N k T {1\over J} {\partial J\over{\partial T}} \quad , \quad
\mu = \mu_{\ast} - k T \ln J -  N k T {1\over J} {\partial
J\over{\partial N}} \quad , \nonumber\\
 U &=& U_{\ast} + NkT^2{1\over J} {\partial
J\over{\partial T}}  \quad , \quad 
  C_V = C^{\ast}_V + 2 N k T {1\over J}
{\partial J\over{\partial T}} + NkT^2 \left[-{1\over{J^2}}  \left(
{\partial J\over{\partial T}} \right)^2 + {1\over J}
{\partial^2J\over{\partial T^2}}\right] \quad .
\end{eqnarray}

\par As can be seen from Eq.(\ref{eq19}),  $J$ depends on the
volume only trough the cut-off whose influence will be seen to be
negligible for the system under study. Thus, the equation of state
$ PV=NkT $ will remain valid, thanks to Eqs.(\ref{eq22}).
The presence of the temperature and
the absence of the number of particles in the expression of $J$
results in the fact that
 the entropy receives two contributions  while the
  last  term of  the chemical potential in Eqs.(\ref{eq22}) vanishes.
  The internal energy  is also modified  and by way of consequence the
specific heat at constant volume too.

The integral giving $J$ can not be computed analytically. However, 
the qualitative features of the theory can be obtained quite easily. When the
deformation parameter goes to zero, $J$  assumes the value one. As
we shall show in a moment, the following parameterization holds:
\be 
\label{eq23} J = 1 + \sigma_1 \frac{T}{T_c}+
\sigma_2 \left( \frac{T}{T_c} \right)^2 + \cdots  \quad . 
\ee
 Working to second order, we find
\be
\label{eq24}
 \frac{T}{J}  \frac{\partial J}{\partial T} =
\sigma_1 \frac{T}{T_c} + \left( 2 
\sigma_2 - \sigma_1^2  \right)
\left( \frac{T}{T_c} \right)^2 
= \log{ \left[ 1 + \sigma_1 \frac{T}{T_c} +
\left( 2  \sigma_2 - \frac{1}{2}
\sigma_1^2 \right) \left( \frac{T}{T_c}
\right)^2 \right]} \quad .
\ee
The last formula's interest lies in the fact that it allows a more
compact expression of the entropy: \be
 \label{eq25}
 S = N k
\left(\frac{5}{2} + \log{\sigma_0} \right) + N k \log{
\left\lbrace \frac{V}{N} \left( \frac{2 \pi m k T}{h^2}
\right)^{3/2} \left[1 + 2 \sigma_1 \frac{T}{T_c}
+ \left( 3  \sigma_2 + \frac{1}{2}
\sigma_1^2 \right) \left( \frac{T}{T_c}
\right)^2 \right] \right\rbrace} \quad , \ee so that an adiabatic
process takes the form: 
\be \label{eq26} V \sim T^{-3/2}
\left[1 - 2 \sigma_1 \frac{T}{T_c} + \left( - 3
 \sigma_2 + \frac{7}{2}
\sigma_1^2 \right) \left( \frac{T}{T_c}
\right)^2 \right] \quad . \ee 
The formulas displayed in
Eqs.({\ref{eq22})} are used to recast  the equation of state and the specific
 heat as
\be 
\label{eq27}
 \rho = P \left[ \frac{3}{2} +
\sigma_1 \frac{T}{T_c} + \left(2 
\sigma_2 - \sigma_1^2 \right) \left(
\frac{T}{T_c} \right)^2 \right] \quad , \quad C_V = N k \left[
\frac{3}{2} + 2 \sigma_1 \frac{T}{T_c} + \left(6
 \sigma_2 - 3 \sigma_1^2
\right) \left( \frac{T}{T_c} \right)^2 \right] \quad . 
\ee
Considering the reaction
\be \label{eq28} a A + b B \rightleftharpoons  d D + e E \quad ,
\ee the expression of the chemical potential $\mu$ shows that the
densities at equilibrium $ X_i = N_i/V$
    obey the modified law of action of masses
\be
\label{eq29}
  \frac{ X_D^d X_E^e}{X_A^a X_B^b} = c^{te} T^{3/2( -a-b+d+e)}
  \left[ 1 + 2 \,d  \, \sigma_{1,D} \,
\frac{T}{T_{c,D}} +2 \, e \, \sigma_{1,E} \,
\frac{T}{T_{c,E}} - 2 \, a \, \sigma_{1,A} \,
\frac{T}{T_{c,A}}- 2 \, b \, \sigma_{1,B} \,
\frac{T}{T_{c,B}} \right] \quad ,
\ee
  since each particle, having its own mass, possesses a specific critical temperature.

\par Let us now evaluate the integral $J$ in order to have numerical
estimates of the constants $\sigma_i$. We shall use 
the Mac Laurin development with a remainder. Let us separate the integral 
of Eq.(\ref{eq19})in two
parts: 
\be
\label{eq30}
 J^{\frac{1}{3}}=I_1 + I_2 =
\frac{1}{\sqrt{\pi}} \left[ \int_0^{\frac{T_c}{2T}} \cdots +
\int_{\frac{T_c}{2T}}^{\omega_{sup}}  \cdots  \right] \quad . \ee
On the second interval, the inequality 
\be \label{eq31}
\frac{1}{1+ \frac{2T}{T_c} \omega} \leq \frac{1}{2} \ee can be
used to obtain the majorization 
\be 
\label{eq32} 
I_2 \leq
\frac{1}{2 \sqrt{\pi}} \int_{\frac{T_c}{2T}}^{\omega_{sup}}
e^{-\omega} \omega^{-1/2} = \frac{1}{2 \sqrt{\pi}}  \left[ \Gamma
\left( \frac{1}{2},\frac{T_c}{2T} \right) - \Gamma \left(
\frac{1}{2},\omega_{sup} \right) \right] \quad . \ee The
asymptotic formula 
\be 
\label{eq33}
 \Gamma(a,z) \sim z^{a-1}
e^{-z} \qquad \mbox{for} \qquad z \rightarrow \infty 
\ee 
shows
that 
\be
\label{eq34} I_2 \leq c^{st}  \left( \frac{T_c}{2T}
\right)^{-1/2} \exp{ \left(-\frac{T_c}{2T} \right)}  + c^{st} \,
\omega_{sup}^{-\frac{1}{2}} \, \exp{(-\omega_{sup})}\quad . \ee
 Since, for a reasonable $\beta$, the characteristic
temperature $T_c$ is reasonably expected to be very high, the integral $I_2$ which contains
the influence of the upper bound can therefore be neglected.

\par To evaluate $I_1$, we shall use the Mac Laurin theorem which
states that for any sufficiently regular function $f$ defined on
an interval $[0,a]$ and for any point $\omega$ on that interval,
there exists another point $\theta(\omega)$ on the same interval
such that 
\be 
\label{eq35}
 f(\omega) = f(0)+f'(0)\omega+
\frac{1}{2}f''(0)\omega^2+\frac{1}{6}f'''( \theta(\omega))\omega^3
\quad .
\ee 
This gives
\be 
\label{eq36}
 \frac{1}{1+2\frac{T}{T_c}\omega} = 1 -
\frac{2T}{T_c} \omega + \left( \frac{2T}{T_c} \right)^2 \omega^2 +
\frac{1}{6} f'''(\theta(\omega)) \omega^3  \quad . \ee
This enables us to find  
\begin{equation}
\label{eq37} \left| | I_1 - \frac{1}{\sqrt{\pi}} \left\lbrace
\int_0^{\frac{T_c}{2T}}d\omega e^{-\omega} \omega^{-1/2} \left[ 1
- \frac{2T}{T_c} \omega + \left( \frac{2T}{T_c} \right)^2 \omega^2
\right] \right\rbrace \right| \leq  \max \vert f'''(x) \vert
\frac{1}{6\sqrt{\pi}} \int_0^{\frac{T_c}{2T}} d\omega e^{-\omega}
\omega^{5/2} \quad .
 \end{equation}
On the interval involved, the following inequality is verified:
\be
 \label{eq38} 
\vert \max{f'''(x)} \vert \leq c^{st} \left( \frac{2T}{T_c}
\right)^3 \quad . \ee 
The integrals remaining in Eq.(\ref{eq37})
can be expressed in terms of complete and incomplete $\Gamma$
functions. The final result reads:
\begin{eqnarray}
\label{eq39} I_1 &=& \frac{1}{\sqrt{\pi}} \left\lbrace \Gamma
\left(\frac{1}{2} \right) - \Gamma \left(\frac{3}{2} \right)
\frac{2T}{T_c} + \Gamma \left(\frac{5}{2} \right) \left(
\frac{2T}{T_c} \right)^2  \right\rbrace
\end{eqnarray}
From this one reads the values of the coefficients $\sigma_i$.

\par  The predictions of usual statistical physics are known to be quite accurate
in ordinary  conditions. We shall use this to give an estimate of
the bound thermodynamics imposes on the deformation parameters.
Consider the Helium   whose specific heat at constant volume
assumes an experimental value comprised between $12.39$ and $
12.41 J K^{-1} mole^{-1}$. The undeformed theory assigns the value
$12.47 J K^{-1} mole^{-1}$ to any non relativistic  gas. Thanks to 
Eqs.(\ref{eq27}), we can write the specific heat in the new theory as
\be 
\label{eq55} C_V = 12.47
\left( 1+ \sigma_1 \frac{T}{T_c} \right) \quad .
\ee
 The
measured value tells us that $ 12.39 < C_V < 12.41$. Assuming $T =
300  °K $, the prediction of the model does not get out of the
experimental bounds provided that $ T_c  \geq  10^6  °K $ which
induces $ \beta \leq 10^{- 45}$. One then finds
 a minimal length uncertainty $ \gamma \leq 10^{-12} $  meters
 which does not
 disagree with
 the bound derived from atomic physics considerations $ \gamma \leq 10^{-16} m $ \cite{LABEL5}
 but is less precise.
 It should be stressed that this only gives an idea of the order
 of magnitude since we did not include interactions between the
 atoms. The estimated temperature  $ T_c $ at which something new should
 happen is too high for the non relativistic  approach to be
 reliable. Thus the interest of the next subsection.

\subsubsection{High Temperatures}

The deformed  Klein-Gordon wave equation  for a massless particle
in a box leads to the spectrum
\be 
\label{eq56} 
E_{\vec n}^2 =
\frac{c^2}{\beta} \left[   \tan^2 {\left(\hbar \sqrt{\beta}
\frac{2 \pi n_x}{L}\right)} +  \tan^2 {\left(\hbar \sqrt{\beta}
\frac{2 \pi n_y}{L}\right)} +  \tan^2 {\left(\hbar \sqrt{\beta}
\frac{2 \pi n_z}{L}\right)} \right] \quad . 
\ee Replacing the sum
by an integral, one obtains for the factor $J$ appearing in the partition 
function the expression
 \be 
 \label{eq57} J= \int_0^{\omega_{sup}} dx
\int_0^{\omega_{sup}} dy  \int_0^{\omega_{sup}} dz  \quad
\varphi(x,y,z) \quad , \ee
 where 
 \be 
 \label{eq58}
\varphi(x,y,z) = \frac{1}{\pi} \left\lbrace \left( 1+
\frac{T^2}{T_{cr}^2} x^2 \right) \left( 1+ \frac{T^2}{T_{cr}^2}
y^2 \right) \left( 1+ \frac{T^2}{T_{cr}^2} z^2 \right) \left(
\exp{ \left( \sqrt{x^2+y^2+z^2} \right)} \right)
\right\rbrace^{-1} \quad , \quad  \omega_{sup} =
\frac{T_{cr}}{T} t \quad . \ee

\par Let us first consider a temperature  verifying $T/T_{cr} < 1$, with an  
upper bound which is practically infinite \mbox{($T_{cr}t/T > 1$, $t$ being given
by Eq.(\ref{eq17}))}. In this case, replacing the domain of integration, 
a cube, by a sphere should not introduce an important error. The result reads
\begin{equation}
\label{eq59}
J = 1 - 12 \frac{{\zeta}(5)}
   {{\zeta}(3)}  \left( \frac{T}{T_{cr}} \right)^2
+288
  \frac{{\zeta}(7)}
   {{\zeta}(3)} \left( \frac{T}{T_{cr}} \right)^4
 \quad .
\end{equation}
From Eqs.(\ref{eq22}), one finds the equation of state and the 
expression of the entropy
\begin{eqnarray}
\label{eq92}
\frac{\rho}{p} &=& 3 - 24 \frac{\zeta(5)}{\zeta(3)} 
\left( \frac{T}{T_{cr}} \right)^2 + 288 \left[ 4 \frac{\zeta(7)}{\zeta(3)} 
 - \left( \frac{\zeta(5)}{\zeta(3)} \right) \right] 
 \left( \frac{T}{T_{cr}} \right)^4  \quad , \nonumber\\
 S &=& 4 N k + N k \log{\left\lbrace 8 \pi \frac{V}{N} \left( \frac{k T}{h c} \right)^3
  \left[ 1 - 36 \frac{\zeta(5)}{\zeta(3)} 
\left( \frac{T}{T_{cr}} \right)^2 + 288 \left( 5 \frac{\zeta(7)}{\zeta(3)} 
 + \left( \frac{\zeta(5)}{\zeta(3)} \right) \right) 
 \left( \frac{T}{T_{cr}} \right)^4  \right]  \right\rbrace} \quad .
\end{eqnarray}

Like in Eqs(\ref{eq25},\ref{eq27}), one obtains small departures from the unmodified 
theory.

\subsubsection{Very high temperatures}

\par What happens at very high temperatures? Like in the
preceding subsection, the most salient features can be captured
from the behavior of $J$. As the  temperature is increased, the
form of the its integrand and its upper bound (Eqs.{(\ref{eq58})}) show 
that J goes to zero. An
approximation of the form 
\be 
\label{eq60} J = \sigma_n \left(
\frac{T_{cr}}{T} \right)^{n} + \sigma_{n+1} \left(
\frac{T_{cr}}{T} \right)^{n+1} + \sigma_{n+2} \left(
\frac{T_{cr}}{T} \right)^{n+2} + \cdots \ee with $n$ a positive
integer will hold. Keeping the
first two corrections one shows, by computations similar to those
of the preceding subsection, that the entropy takes the form 
\be
\label{eq61} 
S = 4 N k + N k \log{ \left\lbrace 8 \pi e^{-n}
\sigma_n \frac{k^3}{(h c)^3} T_{cr}^n \frac{V}{N} T^{3-n} \left[ 1
+ \left( \frac{1}{2} \frac{\sigma_{n+1}^2}{\sigma_n^2} -
\frac{\sigma_{n+2}}{\sigma_n} \right) \left( \frac{T_{cr}}{T}
\right)^2  \right] \right\rbrace } \quad , \ee (from which the
equation of an adiabatic  process can be deduced) while the
equation of state reads 
\be 
\label{eq62} 
\rho = P \left[ (3-n)
- \frac{\sigma_{n+1}}{\sigma_n} \frac{T_{cr}}{T} + \left(
\frac{\sigma_{n+1}^2}{\sigma_{n}^2} - 2
\frac{\sigma_{n+2}}{\sigma_n} \right) \left( \frac{T_{cr}}{T}
\right)^2 \right] \quad . 
\ee The law of action of masses now
assumes the form: 
\be 
\label{eq63}
  \frac{X_D^d X_E^e}{X_A^a X_B^b} = c^{st} \,
T^{(3-n)(d+e-a-b)} \left[ 1 + (d+e-a-b) \left( \frac{1}{2}
\frac{\sigma_{n+1}^2}{\sigma_n^2} -
\frac{\sigma_{n+2}}{\sigma_{n}} \right) \left( \frac{T_{cr}}{T}
\right)^2 \right]\quad . \ee

Let us find the integer $n$. One has
$T/T_{cr} < 1$ and $T_{cr}t/T < 1$. Now, the domain of integration is
small and so the exponential appearing in the function $\varphi$
can be expanded in polynomials. The function $1+T_{cr}^2/T^2 x^2$
admits two Taylor expansions; the region in which $x<T/T_{cr}$
will be denoted $A$; the other one will be denoted $B$. The same
situation occurs for $y$ and $z$; this leads to a partition of the
domain of integration. The most important contribution 
reads
\be
\label{eq64}
 J_{AAA} = \frac{T_{cr}^3}{T^3} \quad  .
\ee
The dependence on the volume is negligible. Comparing with Eq.(\ref{eq60}),
one has $n=3$ so that the equation of state
is, to  first order,
 $\rho \sim 0$. 
  
In this theory, statistics will play a role at high temperature. As is evident
from Eq.(\ref{eq57},\ref{eq58}), $J$ will go to zero as the
temperature increases. The Bose-Einstein distribution 
\be
\label{eq65}
N_i =
\frac{g_i}{  \exp{\left( \frac{\epsilon_i}{kT} - \nu \right)} -1 }
\ee reduces to the Maxwell-Boltzmann's one only in the limiting case
\be 
\label{eq66}
e^{\nu} \ll 1 \quad . \ee
Thanks to the relation giving the total number of particles $N=\sum_i N_i$, one
 is led to the condition 
\be
\label{eq67} 
e^{\nu} = \frac{N}{Z(T,V,1)} \ll 1
\Rightarrow \frac{N}{V} \leq 8 \pi \left( \frac{kT}{hc} \right)^3
J \quad . 
\ee 
One concludes  that neglecting  statistics is
accurate, at very high temperatures, only for systems whose
densities are very small($J \sim T^{-3}$). If this is not the
case, the appropriate integrand in the evaluation of $J$ for
bosons for example
 is 
\be
\label{eq68}
 \varphi(x,y,z) = \frac{1}{\pi_1}\left\lbrace \left( 1+
\frac{T^2}{T_{cr}^2} x^2 \right) \left( 1+ \frac{T^2}{T_{cr}^2}
y^2 \right) \left( 1+ \frac{T^2}{T_{cr}^2} z^2 \right) \left(
\exp{ \left( \sqrt{x^2+y^2+z^2} \right)} -1 \right)
\right\rbrace^{-1} \quad .
\ee
 The normalization constant $\pi_1$
ensures that $J=1$ in the undeformed theory. One finds  the
dominant part is given by
  \be
 \label{eq69}
J_{AAA} = 
    \left(- \frac{\pi }{4} - \log (2\,{\sqrt{2}}) +
      3\,\log (1 + {\sqrt{3}}) \right) \left( \frac{T_{cr}}{T} \right)^2 
      \quad  .
\ee
  In the formula giving $J$, one now has $n=2$ and the associate equation of
   state for a gas of bosons takes the form $\rho = P$. 
   
We treat in more detail the $A_2$ model in the following section. Our choice is
mostly due to the fact that the $A_2$ model, possessing spherical symmetry, gives
$J$ as a one dimensional integral, contrary to the $A_1$ model(see 
Eq.(\ref{eq68})). Apart from leading to simpler formulas, this model seems also
more suitable to the treatment of a Robertson-Walker universe because of
this rotational symmetry.

\subsection{the $A_2$ model}

 In this case, the action of the first position operator on the plane wave
 $\psi_k(t,\vec{\xi}) = \exp{( - i E t + k_1 \xi_1 +  k_2 \xi_2 +  k_3 \xi_3 )}$
  reduces to
\be
\label{eq70}
   \hat p_1 \psi_k(t,\vec \xi) = - i \hbar \psi_k(t,\vec{\xi}) \sum_{r=0}^{\infty} 
    \left( \frac{\hbar^2 \beta}{2} 
{\vec k}^2 \right)^r  \quad .
\ee
The sum of the left hand side converges only when $ \hbar^2 k^2 < 2/\sqrt{\beta}$; this 
is the  cut-off. 
As shown in the last subsection, we do not learn much from the deformed
non relativistic theory; we then go directly to high temperatures.
 The solution to the wave equation gives the dispersion relation
\be
\label{eq71}
E = c \hbar k  \left(1+ \frac{\hbar^2 \beta}{2} 
 k^2 \right)^{-1}  \quad ,
\ee
from which one infers the quantity controlling the departure from the 
unmodified theory, for fermions and bosons:
\be
\label{eq72}
J_{bo} = \frac{1}{2 \zeta(3)} \int_0^{\sqrt{2} \frac{T_{cr}}{T}} dx \,
x^2 \left[ \exp{ \left( \frac{x}{ 1 + \frac{1}{2} \frac{T^2}{T_{cr}^2} x^2 }
 \right) } - 1  \right]^{-1} \quad , \quad
J_{fe} = \frac{2}{3 \zeta(3)} \int_0^{\sqrt{2} \frac{ T_{cr}}{T}} dx \,
x^2 \left[ \exp{ \left( \frac{x}{ 1 + \frac{1}{2} \frac{T^2}{T_{cr}^2} x^2 }
 \right) } + 1  \right]^{-1} \quad .
\ee
For temperatures  smaller than $T_{cr}$, a Taylor expansion of the
term under parentheses and an approximation of the upper bound of the integral by
infinity holds. This leads 
to the following expressions for the equation of state and the entropy 
in the bosonic case:
\begin{eqnarray}
\label{eq73}
\left(\frac{\rho}{p}\right)_{bo} & = & 3 + 60 \frac{\zeta(5)}{\zeta(3)} 
\left( \frac{T}{T_{cr}} \right)^2
+ 360 \left( 21 \frac{\zeta(7)}{\zeta(3)} -
 5 \left( \frac{\zeta(5)}{\zeta(3)} \right)^2 \right) 
\left( \frac{T}{T_{cr}} \right)^4 \quad , \nonumber\\
S_{bo} &=&  4 N k + N k \log{ \left\lbrace 8 \pi \frac{V}{N}
 \left( \frac{k T}{\hbar c} \right)^3  
 \left[1 + 90  \frac{\zeta(5)}{\zeta(3)} 
\left( \frac{T}{T_{cr}} \right)^2
+ 450 \left( 21 \frac{\zeta(7)}{\zeta(3)} + 4
 \left( \frac{\zeta(5)}{\zeta(3)} \right)^2 \right) 
\left( \frac{T}{T_{cr}} \right)^4
  \right] \right\rbrace } \quad ,
\end{eqnarray}

For fermions, the behavior is roughly similar but the details are different:
\begin{eqnarray}
\label{eq74}
\left(\frac{\rho}{p} \right)_{fe} &=& 3 + 75 \frac{\zeta(5)}{\zeta(3)} 
\left( \frac{T}{T_{cr}} \right)^2
+ \frac{45}{2} \left( 441 \frac{\zeta(7)}{\zeta(3)} -
 125 \left( \frac{\zeta(5)}{\zeta(3)} \right)^2 \right) 
\left( \frac{T}{T_{cr}} \right)^4 \quad , \nonumber\\
S_{fe} &= & 4 N k + N k \log{ \left\lbrace 8 \pi \frac{V}{N}
 \left( \frac{k T}{\hbar c} \right)^3  
 \left[1 + \frac{225}{2} \frac{\zeta(5)}{\zeta(3)} 
\left( \frac{T}{T_{cr}} \right)^2
+ \frac{225}{8} \left( 441  \frac{\zeta(7)}{\zeta(3)} +
 100 \left( \frac{\zeta(5)}{\zeta(3)} \right)^2 \right) 
\left( \frac{T}{T_{cr}} \right)^4
  \right] \right\rbrace } \quad  .
\end{eqnarray}

When the temperature becomes of the order of the critical temperature, an 
important change takes place. The domain of integration for $J$ in
 Eq.(\ref{eq72}) becomes small
and so one can approximate the integrand by its  Taylor expansion near 
the origin. The difference between bosons and fermions enters into play through
the difference of signs $\pm$ which leads to different powers in terms of 
the temperature. Developing the full integrand in Eq.(\ref{eq72}) to the fourth
order in $x$, one finds 
\be
\label{eq75}
J_{bo} = \frac{1}{2 \zeta(3)} \left[ \frac{3}{2} \left(\frac{T_{cr}}{T} \right)^2
 - \frac{1}{3} \sqrt{2} \left( \frac{T_{cr}}{T} \right)^3  \right] \quad , \quad
J_{fe} =  \frac{2}{3 \zeta(3)} \left[ \frac{1}{6} \left(\frac{T_{cr}}{T} \right)^3
 - \frac{1}{16} \left( \frac{T_{cr}}{T} \right)^4 \right] \quad .
\ee
A computation to an order greater than the one to which we have limited ourselves
 brings in small corrections to the coefficients $13/16$ , etc. Using 
 Eqs.(\ref{eq62}), one then finds the expression of the entropy and the
equation of state. The difference is significant between the two statistics 
as can be seen from the following equations: 
\be
\label{eq76}
\left(\frac{\rho}{p} \right)_{bo} = 1 + \frac{2}{9} \sqrt{2}  \frac{T_{cr}}{T} +
 \frac{8}{81} \left( \frac{T_{cr}}{T} \right)^2  \quad , \quad
\left( \frac{\rho}{p} \right)_{fe} =  \frac{3}{8}  \frac{T_{cr}}{T} +
 \frac{9}{64} \left( \frac{T_{cr}}{T} \right)^2   \quad .
 \quad
\ee

A numerical computation supports our approximation scheme.
\begin{figure}
\begin{center}
\includegraphics{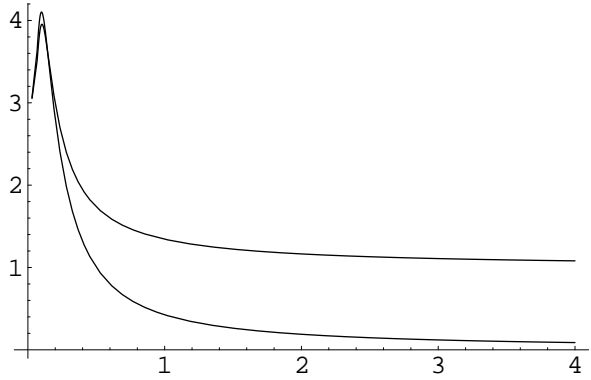}
\end{center}
\caption{$\frac{\rho}{p}$ plotted as a function of $\frac{T}{T_{cr}}$
 for  bosons and fermions}
\end{figure} 
In  Fig.1 we 
plot the ratios between the density and the pressure for a gas of bosons 
(fermions). For temperatures
 smaller than the critical one, these ratios are close to their known value
$3$ in the undeformed theory. The two functions then rise above this value
as predicted by Eqs.(\ref{eq73},\ref{eq74}). They finally tend to the asymptotic values
$1$ and $0$ as obtained in Eqs.(\ref{eq76}).

Let us note that the behavior of the two models, although similar in the 
limiting case of very high temperatures, display some qualitative differences.
As can be seen by comparing Eqs.(\ref{eq92}) and Eqs.(\ref{eq74}), in the $A_1$ 
model the density-pressure ratio does not rise above the usual value $3$, 
contrary to what happens in $A_2$.

We  have here a first manifestation of the domination of bosons in this
context: their density-pressure ratio goes like a constant while the one 
corresponding to fermions vanishes. Their entropy is also the only one
to be considered at scales much higher than $T_{cr}$ as they correspond to 
$n=2$ in Eq.(\ref{eq61}) while for fermions one has 
$n=3$(see Eqs.(\ref{eq75})). Although the ratio we considered is much higher
for bosons, one still can not infer this to be the case for each of its
parts. This will be established for radiation in the next section. 

\section{The black body - radiation}

Let us now analyze how radiation gets affected. 
We shall restrict ourselves to the $A_2$ model for the two reasons explained above.
As the preceding section showed that the equation of state, for example, 
presents the same features at very high temperatures in the two models, we 
expect the same to occur here. 

The formula we will obtain in this section will be  more relevant to the 
very early cosmology. The reason is that at very high temperatures, the
spontaneous symmetry breaking which gives masses to
particles has not taken place yet. One has thus to consider massless particles,
i.e particles with zero chemical potential.

Once again, we follow the notations of \cite{LABEL014}.
The important quantities are
\be
\label{eq77}
q_{bo} = \sum_{\vec l} \log{\left( 1 - \exp{\left(- \frac{\epsilon_{\vec l}}{k T}
 \right)} \right)} = - \log{ Z_{bo}} \quad , \quad
 q_{fe} = - \sum_{\vec l} \log{\left( 1 + \exp{\left(- \frac{\epsilon_{\vec l}}{k T}
 \right)} \right)} = - \log{Z_{fe}} \quad ,
\ee
where $ Z$ is the grand partition function.
The entropy is given by
\be
\label{eq78}
S = - \frac{\partial \Phi}{\partial T} \quad , \quad {\rm with} \quad 
\Phi = k T \log Z \quad , \quad
\ee
while the energy and the particle number read
\be
\label{eq79}
 \quad
U = \sum_{\vec l}  \frac{\epsilon_{\vec l}}{ \exp{\left(
 \frac{\epsilon_{\vec l}}{k T} \right) - 1}} \quad , \quad
N = \sum_{\vec l}  \frac{1}{ \exp{\left(
 \frac{\epsilon_{\vec l}}{k T} \right) - 1}}  \quad .
\ee
It should be noted at this point that considering the $A_1$ model for 
example, if one did not impose an ultraviolet cut off, the periodic dependence
of the energy  (Eq.(\ref{eq56}))would have led to a divergent energy density.

For bosons in the $A_2$ model, the quantity $q$ linked to the partition function by
Eq.(\ref{eq77}) can be written as
\be
\label{eq80}
q = 4 \pi V \left( \frac{k T}{h c}  \right)^3  \quad
\int_0^{\sqrt{2} \frac{T_{cr}}{T} }  dx \,  x^2 
\log{ \left[ 1 - \exp{ \left( - 
\frac{x}{ 1 + \frac{1}{2} \frac{T^2}{T_{cr}^2} x^2 } 
\right)} \right] } \quad ,
\ee
while the energy assumes the following form
\be
\label{eq81}
U = 4 \pi V  \frac{(k T)^4}{(h c)^3}    \quad
\int_0^{\sqrt{2} \frac{T_{cr}}{T} }  dx \, 
\frac{x^3}{ 1 + \frac{1}{2} \frac{T^2}{T_{cr}^2} x^2} 
\left[  \exp{ 
\left(  \frac{x}{ 1 + \frac{1}{2} \frac{T^2}{T_{cr}^2} x^2 } 
\right)} - 1  \right]^{-1} \quad .
\ee
The particle number  admits a similar  integral expression.

For temperatures  greater than or comparable to $T_{cr}$, a Taylor expansion
to fourth order leads to the following expressions:
\begin{eqnarray}
\label{eq82}
p_{bo} &=& \sigma T_c \left[2 - \frac{2}{15} \sqrt{2}
 \frac{ T_c}{T}
 +  \sqrt{2}  \frac{ T}{T_c} \left( \frac{8}{3} \log{\frac{T}{T_c}} +
\frac{112}{45} - \frac{4}{3} \log{2} \right) \right]
\quad , \nonumber\\
 \rho_{bo} &=& \sigma T_c \left[ - 2 +
 \frac{8 }{3} \sqrt{2} \frac{ T}{T_c} +  \frac{4 }{15} \sqrt{2}
 \frac{ T_c}{T} \right] \quad , \quad 
 s_{bo} = \sigma \left[ \frac{2}{15} \sqrt{2} \left( \frac{ T_c}{T} \right)^2 +
   \frac{2}{45} \sqrt{2}   \left( 60 \log{\frac{T}{T_c}} +
116 - 30 \log{2} \right)  \right] \quad ,  \nonumber\\
N_{bo} &=& \frac{\sigma}{k} V  \left[ \frac{1}{3} \frac{T_c}{T} - 
\frac{4}{3} \sqrt{2} + 6 
\frac{T}{T_c}  \right]  \quad , \quad {\rm where} \quad \sigma = 
\pi \frac{k^4 T_c^3}{h^3 c^3} \quad .
\end{eqnarray}
The corresponding quantities for fermions(except the pressure)  are dominated by
constants:
\begin{eqnarray}
\label{eq83}
p_{fe} &=& \sigma T_c \left[ \frac{2}{5} \sqrt{2} 
 \frac{T_c}{T}   - 2 +  \frac{8}{3}  \log{2} \sqrt{2}
 \, \frac{T}{T_c}  \right] \quad , \quad
\rho_{fe} = \sigma T_c  \left(  2 -  \frac{4}{5} \sqrt{2} \,
  \frac{T_c}{T}  \right) \quad , \nonumber\\
s_{fe} &=& \sigma \left[ \frac{8}{3} \sqrt{2} \log{2} - \frac{2}{5} \sqrt{2}
\left( \frac{T_c}{T} \right)^2 \right] \quad , \quad
N_{fe} =   \frac{\sigma}{k} V \left( \frac{4}{3}  \sqrt{2}  -
 \frac{T_c}{T}  \right) \quad .
\end{eqnarray}

The behavior of the energy is depicted in Fig2.
\begin{figure}
\includegraphics{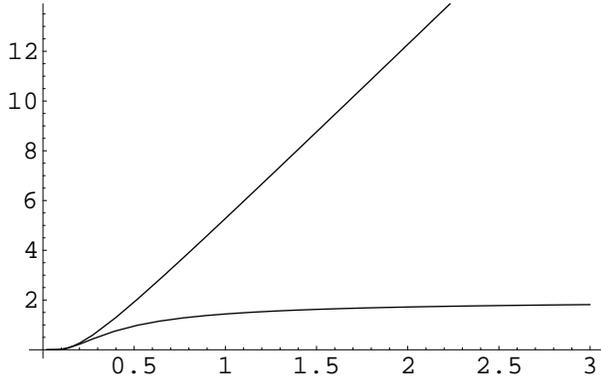}
\caption{The energy densities for fermions and bosons are plotted in 
terms of the temperature. The units adopted are 
  $\sigma T_{cr}$ and $\frac{T}{T_{cr}}$.} 
\end{figure} 
At temperatures below $T_{cr}$, the energy density is polynomial ($ \sim T^4$).
Above $T_{cr}$, it becomes linear as obtained in Eq.(\ref{eq82}) for bosons 
while it goes to a constant for fermions, as shown in Eq.(\ref{eq83}). 

The difference between bosons and fermions in the unmodified theory is encoded 
in the factor $7/8$, for the energy contributions for example. One sees this is 
dramatically changed here. In the usual theory, the ultra relativistic gas
shares the same equation of state with the black body radiation; this feature
is also lost when a minimal uncertainty in length sets in.

\section{Conclusions}

We have studied the thermodynamics induced by a non local theory which 
exhibits a minimal uncertainty in length. We have obtained that a new behavior
sets in at very high temperatures. The difference between fermions and bosons is
more important than in the usual case.

It is worth mentioning some aspects which have not been raised in this work.
At the fundamental level, one can ask if the concept of spin is relevant in 
these theories and, in the case the answer is positive, one still has to study
the relation between  spin and statistics  in the new context.
 As the spin of a
particle is defined, in the modern approach, through the behavior
of its wave function under the Lorentz group, one has to find its generalization
in the new context. For example, taking the ultimate structure of space-time to be given
by a particular noncommutative geometry, the relevant   algebra
  is not the Poincar\'e algebra but its $q$ deformation.
A notion of spin has been defined in these theories and the wave equations
for particles of spin $0,1/2$ and $ 1$  have been found
\cite{LABEL010}. Although the question has not been addressed in
K.M.M. theory, we hope a similar situation to occur. We expect a
generalization  of the notion of spin which  conserves the
spin-statistic theorem.

\underline{Acknowledgments}
 
I warmly thank Ph.Spindel and Ph.de Gottal 
  for useful discussions about   thermodynamical effects in transplanckian 
  physics. I also thank G.Senjanovic, A.Ozpineci, and W.Liao for interesting
  remarks.  

\end{document}